\documentclass[prl,letterpaper,english,reprint,nofootinbib,aps,superscriptaddress,showpacs,showkeys]{revtex4-1}

\usepackage{babel,calc,amsmath,amsthm,amssymb,graphicx,subfigure,xcolor,comment,mathdots}
\usepackage{mathdots}
\usepackage[T1]{fontenc}
\usepackage[unicode=true]{hyperref}
\usepackage{braket}
\usepackage[normalem]{ulem}
\hypersetup{
     colorlinks=true,       	
     linkcolor=blue,          	
     citecolor=red,             
     urlcolor=magenta,          
}

\newif\ifdebug
\debugtrue
\ifdebug
\definecolor{zhliu}{rgb}{0.5, 0.03, 0}

\newcommand{\note}[1]{\textcolor{zhliu}{#1}}
\newcommand\delete{\bgroup\markoverwith{\textcolor{zhliu}{\rule[0.5ex]{2pt}{0.8pt}}}\ULon}

\else

\newcommand{\note}[1]{\ignorespaces}
\newcommand{\delete}[1]{\ignorespaces}

\fi

\begin{document}
\renewcommand{\figurename}{Fig.}
\title{Greenberger-Horne-Zeilinger States: Their Identifications and Robust Violations}

\author{Xing-Yan Fan}
\affiliation{Theoretical Physics Division, Chern Institute of Mathematics, Nankai University,
Tianjin 300071, People's Republic of China}

\author{Jie Zhou}
\affiliation{Theoretical Physics Division, Chern Institute of Mathematics, Nankai University, Tianjin 300071, People's Republic of China}

\author{Hui-Xian Meng}
\affiliation{School of Mathematics and Physics, North China Electric Power University, Beijing 102206, People's Republic of China}

\author{Chunfeng Wu}
\email{chunfeng_wu@sutd.edu.sg}
\affiliation{Science, Mathematics and Technology, Singapore University of Technology and Design, 8 Somapah Road, Singapore 487372, Singapore}

\author{Arun Kumar Pati}
\email{akpati@hri.res.in}
\affiliation{Quantum Information and Computation Group, Harish-Chandra Research Institute, Chhatnag Road, Jhunsi, Allahabad 211 019, India}

\author{Jing-Ling~Chen}
\email{chenjl@nankai.edu.cn}
\affiliation{Theoretical Physics Division, Chern Institute of Mathematics, Nankai University,
Tianjin 300071, People's Republic of China}

\date{\today}

\begin{abstract}
	The $N$-qubit Greenberger-Horne-Zeilinger (GHZ) states are the maximally entangled states of $N$ qubits,
    which have had many important applications in quantum information processing, such as quantum key
    distribution and quantum secret sharing. Thus how to distinguish the GHZ states from other quantum states
    becomes a significant problem. In this work, by presenting a family of the generalized
    Clauser-Horne-Shimony-Holt (CHSH) inequality, we show that the $N$-qubit GHZ states can be indeed
    identified by the maximal violations of the generalized CHSH inequality under some specific measurement
    settings. The generalized CHSH inequality is simple and contains only four correlation functions for any
    $N$-qubit system, thus has the merit of facilitating experimental verification. Furthermore, we present a
    quantum phenomenon of robust violations of the generalized CHSH inequality, in which the maximal
    violation of Bell's inequality can be robust under some specific noises adding to the $N$-qubit GHZ
    states.
\end{abstract}


\maketitle

\textit{Introduction.}---Quantum entanglement is the most surprising nonclassical property of
    composite quantum systems \cite{2009RMPQE} that Schr{\" o}dinger has singled out as ``the characteristic
    trait of quantum mechanics'' \cite{1935Schrodinger}. Nowadays quantum entangled states have become very
    powerful resources for quantum information processing, such as quantum cryptography, quantum
    teleportation, quantum communication protocols, quantum computation, and so on \cite{QCaQI}. Among all
    entangled states, the maximally entangled ones are particularly important for quantum information tasks,
    for instance, Bell's state is the indispensible state for realizing the perfect quantum teleportation~\cite{Tele}.

    As the maximally entangled state of two qubits, Bell's state is usually given by
    $ |{\rm Bell}\rangle = \frac{1}{\sqrt{2}}(|00\rangle+|11\rangle)$. Bell's state has been extensively used
    to reveal quantum nonlocality through quantum violations of some types of Bell's inequalities \cite{BN}.
    The typical two-qubit Bell's inequality is the famous CHSH inequality \cite{1969CHSH}, which is expressed
    as $\mathcal{I}_{\rm CHSH}= A_1B_1+A_1B_2+A_2B_1-A_2B_2 \leq 2$, with $A_i, B_j, (i, j=1, 2)$ being
    measurement settings. For any two-qubit pure state in the Schmidt-decomposition form
    $ |\Psi\rangle=\cos\vartheta |00\rangle+\sin\vartheta |11\rangle$, the maximal violation of the CHSH
    inequality is $\mathcal{I}_{\rm CHSH}^{\rm max}(\vartheta)=2 \sqrt{1+\sin^2(2\vartheta)}$, which
    saturates the well-known Tsirelson's bound $2\sqrt{2}$ \cite{1980Tsirelson} when Bell's state is
    considered. Moreover, by exhausting all states of two qubits, one finds that Bell's state is the only
    state can attain the bound. Therefore, the violations of CHSH inequality have two aspects of
    significance: (i) It may demonstrate the conflict between quantum mechanics and local-hidden-variable
    models \cite{1935EPR,Bell}, and (ii) It can identify the two-qubit maximally entangled state due to its
    maximal quantum violation~\cite{2004PRAZqC}. The latter merit plays a practical role in  experimentally characterizing the entangled state.

    From the viewpoint of maximally entangled state, the GHZ states
    \cite{1989GHZ} are the direct generalizations of Bell's state from bipartite system to multipartite
    system. For the $N$-qubit system, the GHZ state is given by
    \begin{eqnarray}\label{eq:GHZ}
        &&|{\rm GHZ}\rangle = \frac{1}{\sqrt{2}}\biggr(|00\cdots 0\rangle+|11\cdots 1\rangle\biggr),
    \end{eqnarray}
    and its other equivalent forms can be obtained under the local unitary transformations. The importance of
    the GHZ states is considered from two aspects: (i) Theoretically the GHZ states can be used to develop an
    all-verses-nothing proof of Bell's nonlocality for $N$-qubit (i.e., the GHZ theorem), which is a
    significant improvement over Bell's theorem as a way to disprove the concept of ``elements of reality'',
    a concept introduced by the Einstein-Podolsky-Rosen problem in their attempt to prove that quantum theory
    is incomplete, and (ii) GHZ states have many important applications in quantum protocols, such as quantum
    secret sharing. Since about 1990, some researchers have attempted to generalize Bell's inequalities from
    two-qubit to $N$-qubit, the well-known one is the Mermin-Ardehali-Belinski\u{\i}-Klyshko (MABK)
    inequality~\cite{1990PRLM,1992PRAMA,1993PUBK}, which is a tight Bell's inequality for $N$-qubit and
    reduces to the CHSH inequality when $N=2$. In 2004, Chen has explored a rigorous proof that the maximal
    violation of the MABK inequality can be applied to identify the $N$-qubit GHZ states~\cite{2004PRLZqC}. Hence in principle, the MABK inequality offers a useful method for determining the $N$-qubit GHZ state in experiments, however it is not efficient due to the exponentially increasing number of correction-function terms contained in the inequality for $N$ qubits.

    In this Letter, we shall advance the study of the connection between GHZ states and Bell's inequality.
    Our purpose is two-fold: (i) The identification of GHZ states by the maximal violation of Bell's
    inequality is a significant problem with practical interests since experimentally efficient methods are demanded for initializing multi-qubit entangled states required by different quantum protocols. Although the MABK inequality is a possible
    candidate of Bell's inequality to identify GHZ states, its disadvantage in experimental implementation is
    also evident as mentioned earlier. The reason is that the number of correlation-function terms in the MABK inequality
    increases exponentially with the number of qubits ($2^N$ terms for even $N$, or $2^{N-1}$ terms for odd
    $N$), this in turn causes a scalable problem in practical experiments and thus the method based on the MABK inequality is time consuming and resource costly. In this work, we shall overcome
    such a problem by presenting a family of  nontrivial Bell's inequalities, which we call the generalized
    CHSH inequality. The generalized CHSH inequality is tight and simple, and it contains only four
    correlation functions for arbitrary $N$-qubit system, thus facilitating experimental verification.
    (ii) For the GHZ states, we shall show a new quantum phenomenon of robust violations of the generalized
    CHSH inequality. The phenomenon indicates that the maximal violation of Bell's inequality can be robust
    under some specific noises adding to the $N$-qubit GHZ states.

\emph{Identification of GHZ states.}---In order to identify the GHZ states, here we present a family of the
    generalized CHSH inequality for $N$ qubits, which is given by
    \begin{eqnarray}\label{eq:IN}
        \mathcal{I}_N&=& AB+AB'+A'B-A'B'\leq 2,
    \end{eqnarray}
    \begin{eqnarray}\label{eq:E2}
        A=\prod_{j=1}^{N-1} X_j, \ A'=\prod_{j=1}^{N-1} X'_j,\ B=X_N, \ B'=X'_N,
    \end{eqnarray}
    and $X_j, \;X^{\prime}_j$ are variables (valued $\pm 1$) for the $j$-th subsystem classically. Similar to
    the MABK inequality, the generalized CHSH inequality (\ref{eq:IN}) is tight. When $N=2$, the inequality
    (\ref{eq:IN}) naturally reduces to the CHSH inequality, and when $N=3$, it reduces to the third of 46
    Sliwa's inequalities for three qubits~\cite{2003arXivS}.

    Quantum mechanically, we have the measurement operators as
    \begin{eqnarray}\label{eq:E3}
        X_k=\vec{n}_k\cdot \vec{\sigma}_k, \;\;\; X'_k=\vec{n}'_k\cdot \vec{\sigma}_k, \;\;\; (k=1,2,\cdots, N),
    \end{eqnarray}
    where $\vec{\sigma}=({\sigma}_{x},{\sigma}_{y},{\sigma}_{z})$ is the vector of Pauli
    matrices, $\vec{n}_k=(\sin\theta_k\cos\varphi_k,\sin\theta_k\sin\varphi_k,\cos\theta_k)$ and
    $\vec{n}'_k=(\sin\theta'_k\cos\varphi'_k,\sin\theta'_k\sin\varphi'_k,\cos\theta'_k)$ are the measurement
    settings for the $k$-th observer. Accordingly, we have
    \begin{eqnarray}\label{eq:E4}
        &&A=\bigotimes_{j=1}^{N-1} (\vec{n}_j\cdot \vec{\sigma}_j),\
        A'=\bigotimes_{j=1}^{N-1} (\vec{n}'_j\cdot \vec{\sigma}_j),\ \nonumber\\
        &&B=\vec{n}_N\cdot\vec{\sigma}_N,\ B'=\vec{n}'_N\cdot\vec{\sigma}_N.
    \end{eqnarray}

    Now we come to determine the quantum upper bound for the inequality (\ref{eq:IN}). To do this, let us
    define two new observables as
    \begin{eqnarray}\label{eq:E9}
        A''&=&-\dfrac{\mathrm{i}}{2}[A,A']=-\dfrac{\mathrm{i}}{2}(AA'-A'A), \nonumber\\
        B''&=&-\dfrac{\mathrm{i}}{2}[B,B']=(\hat{n}_N\times\hat{n}'_N)\cdot\vec{\sigma}_N.
    \end{eqnarray}
    Let $\alpha_k$ be the angle between two unit vectors $\hat{n}_k$ and $\hat{n}'_k$, then one can obtain
    \begin{equation}\label{eq:ABpp}
        \begin{split}
            & ||A''||\leq |\sin(\alpha_1+\alpha_2+\cdots+\alpha_{N-1})|\leq 1, \\
            & ||B''||=\lVert\hat{n}_N\times\hat{n}'_N\lVert=|\sin\alpha_N|\leq 1,
        \end{split}
    \end{equation}
    iff $\alpha_1+\alpha_2+\cdots+\alpha_{N-1}=\frac{(2\ell+1)\pi}{2}$, $\ell\in \mathbb{Z}$, and
    $\alpha_N=\frac{\pi}{2}$, the equality signs hold. Actually there exist some degrees
    of freedom to select the signs (either $+$ or $-$) before $\alpha_k$, however we can always restrict
    $\alpha_k$ in $[0,\pi]$ and remain the relation $|\sin(\alpha_1+\alpha_2+\cdots+\alpha_{N-1})|=1$
    unchanged in the case of $||A''||=1$. After some calculations, we may have the square operator for the
    Bell function $\mathcal{I}_N$ as
    \begin{eqnarray}\label{eq:E16}
        (\mathcal{I}_N)^2 =4\;\mathbb{I} -[A,A']\otimes[B,B']= 4\bigl(\mathbb{I} +A''\otimes B''\bigr),
    \end{eqnarray}
    here $\mathbb{I}$ is the identity operator. Since
    $||(\mathcal{I}_N)^2||=4\;||\mathbb{I}|| +4\; ||A''|||| B''||\leq 8$, one attains the quantum upper bound
    for the Bell function as $\mathcal{I}_N=\sqrt{||(\mathcal{I}_N)^2||}\leq 2\sqrt{2}$. In other words, the
    maximal violations of the generalized CHSH inequality cannot exceed the Tsirelson's bound.

    Next let us study the quantum violation of inequality (\ref{eq:IN}) for the
    $N$-qubit generalized GHZ states
    $|{\rm GGHZ}\rangle=\cos\vartheta|00\cdots 0\rangle+\sin\vartheta|11\cdots 1\rangle$, whose density
    matrix are expressed as
    \begin{equation}
        \rho_N = \begin{bmatrix}
            \cos^2\vartheta & \cdots & \cos\vartheta\sin\vartheta \\
            \vdots & \ddots & \vdots \\
            \cos\vartheta\sin\theta & \cdots & \sin^2\vartheta
        \end{bmatrix}_{2^N\times2^N},
    \end{equation}
    and except the elements posed on four corners in $\rho_N$, all other elements vanish. For simplicity and
    without loss of generality~\cite{2001JPAVSaGisin}, in this work we always choose the measurement setting
    as $\theta_k=\theta'_k=\pi/2$, which means that the Bloch vectors $\vec{n}_k$ and $\vec{n}'_k$ are
    located on the $xy$-plane. In this way, we have the measurement operators as
    \begin{equation}\label{eq:Xjk}
        X_k=\begin{bmatrix}
            0 & e^{-\mathrm{i}\varphi_k} \\
            e^{\mathrm{i}\varphi_k} & 0
        \end{bmatrix},\;\;\;
        X'_k=\begin{bmatrix}
            0 & e^{-\mathrm{i}\varphi'_k} \\
            e^{\mathrm{i}\varphi'_k} & 0
        \end{bmatrix}.
    \end{equation}
    We then have
    \begin{eqnarray}\label{eq:MoIN}
        \langle\mathcal{I}_N\rangle&=&\text{tr}(\rho_N\mathcal{I}_N) = \sin2\vartheta\;
            [\cos(a+b)+\cos(a+b')\nonumber\\
        && +\cos(a'+b)-\cos(a'+b')],
    \end{eqnarray}
    with
    \begin{equation}
            a=\displaystyle\sum_{j=1}^{N-1}\varphi_j,\quad b=\varphi_N, \quad
            a'=\displaystyle\sum_{j=1}^{N-1}\varphi'_j,\quad b'=\varphi'_N.
    \end{equation}
    It can be verified that
    \begin{eqnarray}\label{condit}
        &&\cos(a+b)+\cos(a+b')+\cos(a'+b)-\cos(a'+b') \nonumber\\
        &&= 2 F\cos\Bigl(\dfrac{b-b'}{2}\Bigr) -  2 G\sin\Bigl(\dfrac{b-b'}{2}\Bigr) \nonumber\\
        &&=2\sqrt{F^2+G^2} \cos\Bigl(\dfrac{b-b'}{2}+\delta\Bigr)\leq 2\sqrt{2},
    \end{eqnarray}
    with $F=\cos\Bigl(a+\frac{b+b'}{2}\Bigr)$, $G=\sin\Bigl(a'+\frac{b+b'}{2}\Bigr)$, and $\tan\delta=G/F$.
    The equal sign in Eq. (\ref{condit}) occurs for $F^2=G^2=1$ and $\cos\Bigl(\frac{b-b'}{2}+\delta\Bigr)=1$,
    which leads to the following conditions:
    \begin{eqnarray}\label{condit2}
        && a+\frac{b+b'}{2}=p\pi, \;\; a'+\frac{b+b'}{2}=\frac{(2q+1)\pi}{2},\nonumber\\
        && \frac{b-b'}{2}+\delta=2r \pi,\;\; \tan\delta=\pm 1.
    \end{eqnarray}
    with $p, q, r\in\mathbb{Z}$. Note that the last two equations in Eq. \eqref{condit2} is
    same as the constraint of $\lVert B''\lVert=1$. From above equations, we have
    \begin{equation}\label{eq:MVg}
        \langle\mathcal{I}_N(\vartheta)\rangle_{\text{max}}=2\sqrt{2}\sin(2\vartheta),
    \end{equation}
    which implies that the GHZ states can reach the Tsirelson's bound when we take $\vartheta=\pi/4$.

    By choosing some appropriate measurement settings, now we can identify the $N$-qubit GHZ states by maximal
    violations of the general CHSH inequality. We have the following theorem:

    \emph{Theorem 1.}---Under the conditions of measurement settings $\varphi_j=0$,
        $\varphi'_j=\frac{\pi}{2(N-1)}$, $(j=1,2,...,N-1)$, and $\varphi_N=-\frac{\pi}{4}$,
        $\varphi'_N=\frac{\pi}{4}$, a $N$-qubit state $|\psi\rangle$ satisfies
        \begin{equation}
            \langle\mathcal{I}_N\rangle= \bigr\langle\psi\bigr|\mathcal{I}_N\bigl|\psi\bigr\rangle=2\sqrt{2},
        \end{equation}
        iff it is the $N$-qubit GHZ state in the form of Eq. (\ref{eq:GHZ}).

        \emph{Proof.}---Firstly, we need to prove that if $|\psi\rangle$ is the $N$-qubit GHZ state, then one
            has $\langle\mathcal{I}_N\rangle = 2\sqrt{2}$. Based on the conditions in \emph{Theorem 1}, we
            have $a=\sum_{j=1}^{N-1}\varphi_j=0$, $a'=\sum_{j=1}^{N-1}\varphi'_j=\frac{\pi}{2}$,
            $b=-\frac{\pi}{4}$, $b'=\frac{\pi}{4}$, this yields
            $ a+\frac{b+b'}{2}=0$, $a'+\frac{b+b'}{2}=\frac{\pi}{2}$, $\delta=\frac{\pi}{4}$,
            $\frac{b-b'}{2}+\delta=0$, which satisfy the conditions in Eq. (\ref{condit2}). Therefore, it is
            easy to prove that if $|\psi\rangle=|{\rm GHZ}\rangle$, then
            $\langle\mathcal{I}_N\rangle= 2\sqrt{2}$ holds.

            Secondly, we need to prove that, under the conditions in \emph{Theorem 1}, the GHZ state is the
            unique state that can satisfy $\langle\mathcal{I}_N\rangle= 2\sqrt{2}$. To reach this purpose,
            one need to show the largest eigenvalue of Bell-function matrix is not degenerate. Under the
            constraints in \emph{Theorem 1},  one obtains
            \begin{equation}
                X_j=
                \begin{bmatrix}
                    0 & 1 \\
                    1 & 0
                \end{bmatrix},\
                X'_j=
                \begin{bmatrix}
                    0 & e^{-\mathrm{i}\varphi'_j} \\
                    e^{\mathrm{i}\varphi'_j} & 0
                \end{bmatrix},
                \end{equation}
            with $j=1,2,...,N-1$, and
            \begin{equation}
                B=X_N=
                \begin{bmatrix}
                    0 & e^{\mathrm{i}\frac{\pi}{4}} \\
                    e^{-\mathrm{i}\frac{\pi}{4}} & 0
                \end{bmatrix},\
                B'=X'_N=
                \begin{bmatrix}
                    0 & e^{-\mathrm{i}\frac{\pi}{4}} \\
                    e^{\mathrm{i}\frac{\pi}{4}} & 0
                \end{bmatrix}.
            \end{equation}
            Then,
            \begin{equation}
                \begin{split}
                    A &=\text{C-diag}\left(1,1,...,1,1\right), \\
                    A'&= \text{C-diag}\left(
                        -\mathrm{i},e^{-\mathrm{i}\frac{(N-3)\pi}{2(N-1)}},...,
                            e^{\mathrm{i}\frac{(N-3)\pi}{2(N-1)}},\mathrm{i}\right),
                \end{split}
            \end{equation}
            where C-diag($e_1,e_2,...,e_{N-2},e_{N-1}$) represents the counter-diagonal matrix with $e_j$ the
            nonzero elements in the $j$-th row. After that, we deduce the Bell-function matrix as
            \begin{eqnarray}
                \mathcal{I}_N &=& \text{C-diag}\biggl(
                    2\sqrt{2},0,\sqrt{2}\left(1+e^{\mathrm{i}\frac{\pi}{N-1}}\right),
                        \sqrt{2}\left(1-e^{\mathrm{i}\frac{\pi}{N-1}}\right),..., \nonumber\\
                    && \sqrt{2}(1+e^{\mathrm{i}\frac{(N-2)\pi}{N-1}}),
                        \sqrt{2}(1-e^{\mathrm{i}\frac{(N-2)\pi}{N-1}}),0,2\sqrt{2}\biggr).
            \end{eqnarray}
            Obviously, the largest eigenvalue $2\sqrt{2}$ of matrix $ \mathcal{I}_N$ is unique. This ends the
            proof.

\emph{The Phenomenon of Robust Violations of Bell's Inequality.}---Usually the maximal violations of Bell's
    inequality are not robust. Let us take the CHSH inequality as an example. (i) Pure state case. As
    mentioned above, for Bell's state, the maximal violation is
    $\mathcal{I}_{\rm CHSH}^{\rm max}(|{\rm Bell}\rangle)=2 \sqrt{2}$. It is well-known that for any
    two-qubit pure state $|\psi\rangle$, the maximal violation is  given by
    $\mathcal{I}_{\rm CHSH}^{\rm max}(|\psi\rangle)=2 \sqrt{1+\mathcal{C}^2}$, where $\mathcal{C}\leq 1$ is
    the entanglement degree of the state $|\psi\rangle$. For a pure state generated from the
    superposition of Bell's state and an arbitrary two-qubit pure state $|\Phi\rangle$, i.e.,
    $|\Psi\rangle=\sqrt{1-\epsilon^2}|{\rm Bell}\rangle+\epsilon |\Phi\rangle$, its maximal violation
    will be always less than $2\sqrt{2}$ because $\mathcal{C}(|\Psi\rangle)<1$ for any $\epsilon >0$. (ii)
    Mixed state case. Let us consider a two-qubit mixed states
    $\rho=(1-\epsilon)\rho_{\rm Bell}+\epsilon \rho_{\rm noise}$ (with $0\leq \epsilon \leq 1$), which is a
    convex sum of density matrix of Bell's state and a certain noise. In this case, one also finds that the
    maximal violation cannot reach $2\sqrt{2}$. For instance, let us take $\rho_{\rm noise}=\mathbb{I}/4$
    that represents the white noise, and $\rho$ becomes the well-known Werner state \cite{1989PRAWerner}, we
    then have the quantum violation as
    $\mathcal{I}_{\rm CHSH}^{\rm max}(\rho)=(1-\epsilon)2\sqrt{2}<2\sqrt{2}$. Therefore, any small
    disturbance adding to Bell's state will let the maximal violation deviate from the Tsirelson's bound.

    In this work, we present for the first time the quantum phenomenon of robust violations of Bell's
    inequality based on the generalized CHSH inequality (\ref{eq:IN}). For $N=2$, there is not the phenomenon
    of robust violations, as we have analyzed above.  For $N=3$, there are only trivial robust violations,
    because for at least one of observers, his two measurement settings are the same or opposite (i.e.,
    $\vec{n}_j=\vec{n}'_j$ or $\vec{n}_j=-\vec{n}'_j$). For $N\geq 4$, there are nontrivial phenomena of
    robust violations of Bell's inequality, in this case each observer performs two distinct measurements on
    his qubit. In the following, to address this topic, we will provide an explicit example for $N=4$.

\emph{Example.}---Let us take $\varphi_1=\varphi_2=\varphi'_4=0$,
    $\varphi'_1=\varphi'_2=\varphi_4=\frac{\pi}{2}$,
    $\varphi_3=-\frac{\pi}{4}$, and $\varphi'_3=\frac{\pi}{4}$, namely, in the $xy$-plane, the measurement
    directions are $\vec{n}_1=\vec{n}_2=\vec{n}'_4=\hat{x}$, $\vec{n}'_1=\vec{n}'_2=\vec{n}_4=\hat{y}$,
    $\vec{n}_3=(\hat{x}-\hat{y})/\sqrt{2}$, $\vec{n}'_3=(\hat{x}+\hat{y})/\sqrt{2}$. we  have
    $a=\sum_{j=1}^3\varphi_j=-\frac{\pi}{4}$, $a'=\sum_{j=1}^3\varphi'_j=\frac{5\pi}{4}$,
    $b=\frac{\pi}{2}$, $b'=0$, this yields $ a+\frac{b+b'}{2}=0$, $a'+\frac{b+b'}{2}=\frac{3\pi}{2}$,
    $\delta=-\frac{\pi}{4}$, $\frac{b-b'}{2}+\delta=0$, which satisfy the conditions in Eq. (\ref{condit2}).
    Thus for the four-qubit GHZ state $|{\rm GHZ}_0\rangle = \frac{1}{\sqrt{2}}(|0000\rangle+|1111\rangle)$,
    one gets $\mathcal{I}_{N=4}^{\rm max}= 2\sqrt{2}$.

    Due to the above measurement settings, one then has
    \begin{eqnarray}
            && A = \text{C-diag}(\mu,\nu,\mu,\nu,\mu,\nu,\mu,\nu) \nonumber\\
                &&    A' =\text{C-diag}(-\nu,-\mu,\nu,\mu,\nu,\mu,-\nu,-\mu), \nonumber\\
                &&  B = \text{C-diag}\left(-\mathrm{i},\mathrm{i}\right),\nonumber\\
                    &&  B'=\text{C-diag}\left(1,1\right).
    \end{eqnarray}
    with $\mu=(1+\mathrm{i})/\sqrt{2}$ and $\nu=(1-\mathrm{i})/\sqrt{2}$. This leads to
    \begin{eqnarray}
                    \mathcal{I}_{4} &=& \text{C-diag}\bigl(2\sqrt{2},0,0,2\sqrt{2},0,\mathrm{i}2\sqrt{2},
                        -\mathrm{i}2\sqrt{2},0, \nonumber\\
                            &&\quad 0,\mathrm{i}2\sqrt{2},-\mathrm{i}2\sqrt{2},0,2\sqrt{2},0,0,2\sqrt{2}\bigr).
    \end{eqnarray}
    For the matrix of $\mathcal{I}_{4}$, the largest eigenvalue $2\sqrt{2}$ is four-fold degeneracy, and the
    corresponding orthonormal eigenstates are all GHZ states in different forms:
    \begin{eqnarray}
                & & |{\rm GHZ}_0\rangle =\dfrac{1}{\sqrt{2}}\Bigl(|0000\rangle+|1111\rangle\Bigr), \nonumber\\
                & & |{\rm GHZ}_1\rangle =\dfrac{1}{\sqrt{2}}\Bigl(|0011\rangle+|1100\rangle\Bigr), \nonumber\\
                & & |{\rm GHZ}_2\rangle =\dfrac{1}{\sqrt{2}}\Bigl(\mathrm{i}|0101\rangle+|1010\rangle  \Bigr), \nonumber\\
                & & |{\rm GHZ}_3\rangle =\dfrac{1}{\sqrt{2}}\Bigl(-\mathrm{i}|0110\rangle+|1001\rangle  \Bigr).
    \end{eqnarray}
    It is easy to verify that for the following pure state
    \begin{eqnarray}
        |\psi\rangle=\sqrt{1-\sum_{j=1}^3\epsilon_j^2}\;
            |{\rm GHZ}_0\rangle + \sum_{j=1}^3 \epsilon_j |{\rm GHZ}_j\rangle,
    \end{eqnarray}
    and mixed state
    \begin{eqnarray}
        \rho=(1-\sum_{j=1}^3\epsilon_j)\; \rho_0+ \sum_{j=1}^3 \epsilon_j \rho_j,
    \end{eqnarray}
    with $\rho_k=|{\rm GHZ}_k\rangle\langle {\rm GHZ}_k|$, and for any $0<\epsilon_j<1, (j=1, 2, 3)$,
    one always has the maximal violations as $\mathcal{I}_{4}^{\rm max}= 2\sqrt{2}$, thus demonstrating the
    quantum  phenomenon of robust violations of Bell's inequality for the GHZ state.

\emph{Conclusion.}---In summary, we have proposed a family of generalized CHSH inequality for $N$ qubits, and
    based on which we have also addressed two significant topics related to the connection between the
    $N$-qubit GHZ states and Bell's inequality. The first one is that we have successfully identified the
    $N$-qubit GHZ states by the maximal violation of the generalized CHSH inequality. The generalized CHSH
    inequalities are tight, indeed we have checked the tightness of the inequality for $N\leq 7$ and believe
    that it is also true for larger $N$. Moveover, the generalized inequality is very simple in the sense
    that it contains only four correlation-function terms for arbitrary $N$, thus providing a friendly way
    for experimental implementations with reduced complexity. In various quantum protocols with entangled states as computational resource, efficient characterization of the entangled states is of essential importance in the initial step of implementing the protocols. However for multi-qubit entangled states, it is still challenging to verify the effectiveness of the initialization of the states even though some efforts have been taken in the literature \cite{2018PRLMontanaro, 2018PRXMorimae, 2019PRapplZhang, 2019PRapplHayashi, 2020PRLGuo}. Our generalized inequality is a timely contribution to provide an efficient verification method for the $N$-qubit GHZ state in experiments. The second one is that for the GHZ states, we
    have present a quantum phenomenon of robust violations of Bell's inequality. We find that for the GHZ
    states, the maximal violations of the generalized CHSH inequality are robust under some specific noises
    adding to the $N$-qubit GHZ states. We anticipate further experimental progresses in this direction in
    the near future.

\textit{Acknowledgements.}---J.L.C. was supported by National Natural Science Foundations of China
    (Grant Nos. 11875167 and 12075001). X.Y.F was supported by Nankai Zhide Foundation. H.X.M. was supported
    by the National Natural Science Foundations of China (Grant No. 11901317).

\end{document}